%% file: main.tex
\documentclass[10pt, conference, letterpaper]{IEEEtran}

\usepackage{listings}
\usepackage{amsfonts}

\usepackage{multicol}

\usepackage{blindtext}
\usepackage{amsfonts}
\usepackage{amsmath, amsthm, amssymb}
\usepackage{times}
\usepackage{url}
\usepackage{verbatim}
\usepackage{array}
\usepackage{mathrsfs}
\usepackage{enumitem}
\usepackage{textcomp}

\newcommand{\subparagraph}{}
\usepackage[compact]{titlesec}
\usepackage[ruled, lined, linesnumbered]{algorithm2e}
\SetAlFnt{\small}

\SetCommentSty{mycommfont}

\usepackage{amsfonts}
\usepackage{bbm}
\usepackage{soul}
\usepackage{color}
\usepackage{graphics}
\usepackage{graphicx,epsfig}
\usepackage{lipsum,graphicx,subcaption}
\graphicspath{{figures/}}
\usepackage{caption}
\usepackage{subcaption}

\usepackage{cleveref}
\usepackage{wrapfig}
\usepackage{url}
\usepackage{url}
\usepackage{amsmath,amssymb,epsfig,amsthm,psfrag,epsf, multirow,wrapfig, graphicx}
\usepackage[english]{babel}
\usepackage{epstopdf}
\usepackage{float}
\usepackage{color}
\usepackage[table,xcdraw]{xcolor}
\usepackage{amsmath}
\usepackage{amsfonts}
\usepackage{amssymb}
\usepackage{mathtools}

\usepackage{tabularx}
\usepackage{adjustbox}
\usepackage{physics}

\usepackage{amsfonts}
\usepackage{amssymb}
\usepackage{caption}
\usepackage{subcaption}
\usepackage{tabularx}
\usepackage{comment}
\usepackage[mathscr]{euscript}
\usepackage{amsbsy}
\usepackage{stmaryrd}
\usepackage{mathtools}
\usepackage[utf8]{inputenc}
\graphicspath{{figs/}}

\usepackage{booktabs}
\usepackage{graphicx,epsfig}
\usepackage{graphics}
\usepackage{multirow}
\usepackage{rotating}
\usepackage{caption}
\usepackage{subcaption}
\usepackage{amsmath}
\usepackage{amssymb}
\usepackage{makecell}
\usepackage{mathtools}
\usepackage{comment}
\usepackage{multirow}

\usepackage{footnote}
\usepackage{tablefootnote}
\usepackage{scalerel}

\usepackage{tabularx}
\usepackage{mdframed}
\usepackage{lipsum}
\makeatletter
\newcommand*{\rom}[1]{\expandafter\@slowromancap\romannumeral #1@}
\makeatother
\usepackage{nccmath}

\usepackage{relsize}

\renewcommand\IEEEkeywordsname{Keywords}

\DeclareMathSymbol{\shortminus}{\mathbin}{AMSa}{"39}

\newcommand{\abbrev}{\texttt{HOGMT}}

\allowdisplaybreaks
\begin{document}
\def\eg{\mbox{\em e.g.}, }


\title{Waveforms for xG Non-stationary Channels}

\author{
\IEEEauthorblockN{Zhibin Zou, and Aveek Dutta}
    \IEEEauthorblockA{Department of Electrical and Computer Engineering\\
    University at Albany SUNY, Albany, NY 12222 USA\\
    \{zzou2, adutta\}@albany.edu}
    \vspace{-5ex}

}
    
\maketitle

\input{abstract}

\input{intro}
\input{background}
\input{decomposition}
\input{precoding}
\input{modulation}
\input{futurework}

\input{conclusion}





\input{ack}

\bibliographystyle{IEEEtran}
\bibliography{references}

\input{bio}



\end{document}

%% file: abstract.tex
\begin{abstract}
Waveform design 
aims to achieve orthogonality among data signals/symbols across all available Degrees of Freedom (DoF) to avoid interference while 
transmitted over the channel. 
In general, precoding decomposes the channel matrix into desirable components in order to 
construct a precoding matrix, which is combined with 
the data signal to 
orthogonality in the spatial dimension. On the other hand, modulation uses orthogonal carriers in a certain signal space to carry data symbols
with minimal interference from other symbols. 
However, it is widely evident that next Generation (xG) wireless systems will experience very high mobility, density and time-varying multi-path propagation that will 
result in a highly non-stationary of the channel states.
Conventional precoding methods using SVD or QR decomposition, are unable to capture these joint spatio-temporal variations as those techniques treat the space-time-varying channel as separate independent spatial channel matrices and hence fail to achieve joint spatio-temporal orthogonality. Meanwhile, the carriers in OFDM and OTFS modulations are unable to maintain the orthogonality in the frequency and delay-Doppler domain respectively, due to the higher order physical variation like velocity (Doppler effect) or acceleration (time-varying Doppler effect). In this article, we review a recent method called High Order Generalized Mercer's Theorem ({\abbrev}) for orthogonal decomposition of higher dimensional, non-stationary channels and its application to MU-MIMO precoding and modulation. We conclude by identifying some practical challenges and the future directions for waveform design for MU-MIMO non-stationary channels based on {\abbrev}. 

\end{abstract}

%% file: intro.tex
\section{Introduction}
\label{sec:intro}

\begin{figure}
    \centering
    \includegraphics[width=\linewidth]{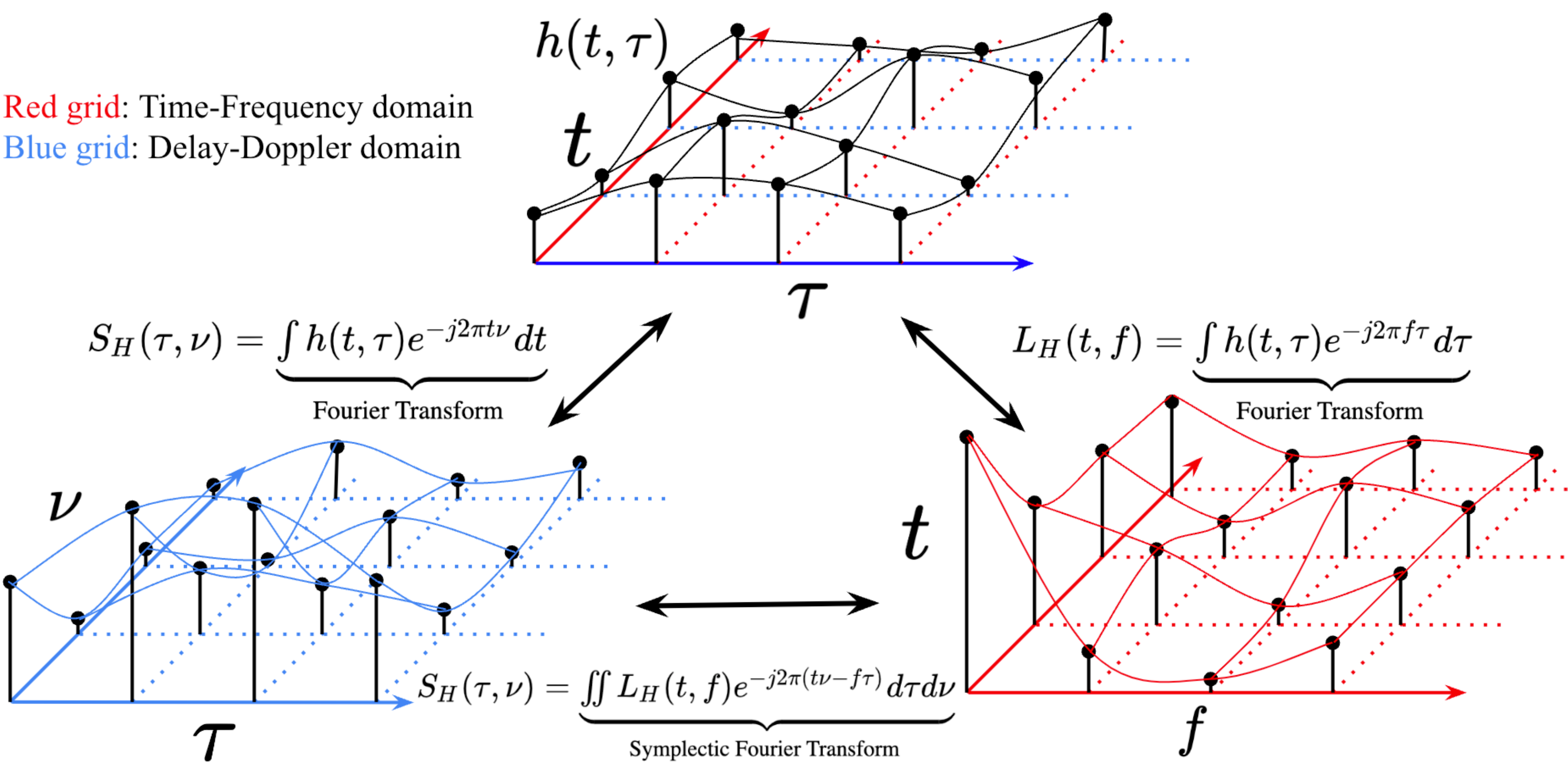}
    \caption{General Linear Time Varying model transition in 4-D domain (time, frequency, delay and Doppler).}
    \label{fig:4-D_relation}
    \vspace{-10pt}
\end{figure}

Any general Linear Time Varying (LTV) channel can be represented in different domains by using different representations that can be used to design orthogonal waveforms. Figure ~\ref{fig:4-D_relation} shows three such domains; a) Time-Delay Domain, which is defined by the \textit{time-varying impulse response} $h(t,\tau)$; b) Delay-Doppler domain, represented by the \textit{spreading function} $S_H(\tau,\nu)$ and; c) Time-Frequency domain, defined by the \textit{Time-Frequency (TF) transfer function}, $L_H(t,f)$. These domains are inter-related either by Fourier Transform (FT) or Symplectic Fourier Transform (SFT)~\cite{MATZ20111} as shown in figure~\ref{fig:4-D_relation}. 
Historically, waveforms have been designed based on the perspective of the channel presentation. For example, Orthogonal Frequency Division Multiplexing (OFDM) modulation uses orthogonal carriers in the frequency domain, which also alleviates Inter-symbol Interference (ISI) caused by multiple path delays. However, OFDM symbols cannot maintain its orthogonality when the channel is affected by relative difference of transceiver velocities, which is also known as Doppler effect. This leads to Inter-Carrier Interference (ICI). Recently, Orthogonal Time Frequency Space (OTFS) modulation has been proposed~\cite{ OTFS_2018_Paper}, 
which design carriers in the delay-Doppler domain. The OTFS symbols achieve joint orthogonality in the time-frequency domain so as to simultaneously cancel both ISI and ICI. 

From another perspective, precoding is widely adopted to exploit the spatial degree of freedom (DoF) but requires the Channel State Information (CSI) at the transmitter. SVD-based precoding for MIMO channels is able to decompose the MIMO channel into parallel subchannels and achieves capacity via the \textit{water-filling algorithm}. Alternately, Dirty Paper Coding (DPC) for MU-MIMO channels~\cite{Cho2010MIMObook} can preemptively cancel Inter-User Interference (IUI) at the transmitter. With perfect CSI, DPC is proven to achieve downlink MU-MIMO channel capacity~\cite{LeeDPC2007}. However, these waveform design methods assume that the channel is a stationary process. The non-stationary (NS) channel model is discussed in~\cite{Matz2005NS}, where the second-order statistics of the channel is varying across a 4-dimensional domain consisting of time-frequency and delay-Doppler. For MU-MIMO non-stationary channels, the second-order derivative of the path difference, acceleration difference (time-varying Doppler effect) will also cause Inter-Doppler Interference (IDI), meaning the OTFS symbols can not maintain their orthogonality in this case. Meanwhile, as the channel is rapidly time-varying, it is modeled as the channel matrix that is also a function of time, resulting in a spatio-temporal channel tensor. In such cases, the SVD-based precoding and DPC will treat the channel as separate independent spatial channel matrices, which is unable to capture the joint spatio-temporal variation of the channel and thus the decomposed subchannels can not be jointly orthogonal across all DoF.

\textit{``How to decompose non-stationary channels and what is the corresponding signaling scheme"} is a challenging open problem in the literature~\cite{2006MatzOP}. To address this, High Order Generalized Mercer's Theorem ({\abbrev})~\cite{ZouICC22} has been shown to be an effective framework for decomposing MU-MIMO non-stationary channel. With proper knowledge of the CSI, it can decompose the high-dimensional channels into independent subchannels along each DoF. In this article, we discuss the possibilities of a novel waveform based on {\abbrev} called Multidimensional Eigenwave Multiplexing (MEM) modulation and a precoding technique called {\abbrev}-Precoding~\cite{ZouICC22, zzou2023Tcom}, that are particularly suited for the for xG non-stationary channels.

The article is organized as follows: We begin by highlighting the challenges in designing waveforms for MU-MIMO non-stationary channels and the limitations of conventional methods. This is followed by a thorough review of {\abbrev} and a comparative study with other decomposition methods. Next, we discuss the novel {\abbrev}-Precoding snd MEM modulation with comparison to SoTA techniques. Then a qualitative comparison between the two techniques followed by potential future research directions. Finally, we conclude the article.  

%% file: background.tex
\section{Challenges of Precoding and Modulation for MU-MIMO NS channels}
\label{sec:back}
\subsection{Non-stationarity in Wireless Channels}

\begin{table}
\centering
    \def\arraystretch{1}
    \setlength{\tabcolsep}{.2em}
\begin{tabular}{|p{2.4cm}|p{3.75cm}|p{2.3cm}|}
\hline
\multicolumn{1}{|c|}{\textbf{Applications}} & \multicolumn{1}{c|}{\textbf{Source of NS}} &  \multicolumn{1}{c|}{\textbf{Performance Gap}} \\ \hline
V2X, HST, UAV, Massive/XL-MIMO, RIS, VLC, THz, Underwater & High mobility, weather, time-varying blockage and scatterers, UAV attitude, spatial visibility, reflection angles and atmospheric absorption.   &  \multirow{4}{2.2cm}[0em] {\textbf{BER with SoTA:} ${>}10^{-1}$ at 20dB SNR for BPSK~\cite{Jaime2020V2X}
  
  \textbf{Target Metrics:} $10^{-7}{<}\textbf{BER}{<}10^{-5}$ 
   \newline 
  \newline } \\ \hline
\end{tabular}
\caption{Overview of applications in NS channels: Impairments, SoTA and performance gap.}
\vspace{-10pt}
\label{tab:ns}
\end{table}

The entire apparatus for error-free communication is centered on models and analysis of the Probability Distribution Function (PDF) that are translated to implementable transceiver algorithms. Now, symbol detection is relatively tractable for linear, stationary channels with known distributions by employing the gamut of tools for Bayesian inference.  However, there are many instances, such as next Generation (xG), where the PDF of the channel is either non-linear, high-dimensional, or statistically non-stationary. Depending on the features of the channel, non-stationarity may arise from the time-varying Doppler in Vehicle-to-everything (V2X), High-speed Train (HST) and Unmanned Aerial Vehicle (UAV) channels, from time-varying multipath in MU-MIMO channels and due to time-varying blockage in Reconfigurable Intelligent Surfaces (RIS), Terahertz (THz) links, Visible Light Communications (VLC) and underwater channels. Non-stationarity of the channel 
is often measured in terms of the stationarity interval (SI)  
in time, frequency or space. Unfortunately, conventional methods applied to Non-stationary (NS) channels are only able to achieve modest error rates that is inadequate to support high data rate wireless applications like mobile AR/VR/XR, aerial communications and 4K/8K HDR video streaming services~\cite{wang2018survey}. These will require waveforms that are capable of achieving orders of magnitude improvement in Bit Error Rate (BER) across all types of channels and applications. Table~\ref{tab:ns} provides a summary of the literature about this challenging and less explored area of NS channels.

\subsection{Conventional Precoding and Limitations}
Precoding has been widely investigated for stationary channels, 
where the orthogonality along each DoF 
is enforced by decomposing them using linear algebraic tools (\eg singular value decomposition (SVD) or QR decomposition \cite{Cho2010MIMObook}) 
leading to capacity achieving strategies 
typically under the block fading assumption, which may not be possible in NS channels. Therefore, non-stationarity of the channel 
, 
leads to \textit{catastrophic} error rates even with state-of-the-art precoding \cite{AliNS0219} 
because these are optimized for stationary channels do not ensure interference-free communication when the distribution changes rapidly over time. 
Although, DPC yields interference-free communication with perfect CSI, current implementation by QR decomposition of independent channel matrices ($\mathbf{H}$), does not capture the joint variations across multiple domains (space (users/ antennas), time-frequency or delay-Doppler).
This joint interference renders conventional decomposition techniques incapable of achieving flat-fading. Hence, the new decomposition method is critically important to obtain jointly orthogonal components in high-dimensional, time-varying processes like NS channels. Furthermore, since any channel can be generated as a special case of a general non-stationary channel, precoding for NS channels will also apply to any and all types of wireless channel \cite{MATZ20111}). This is the main motivation of this article, which reviews the novel {\abbrev} decomposition and precoding that extracts orthogonal eigenfunctions from NS channels to mitigate joint space-time interference, which is a significant addition to the literature.

\subsection{Conventional Modulation and Limitations}
\begin{figure}
    \centering
    \includegraphics[width=\linewidth]{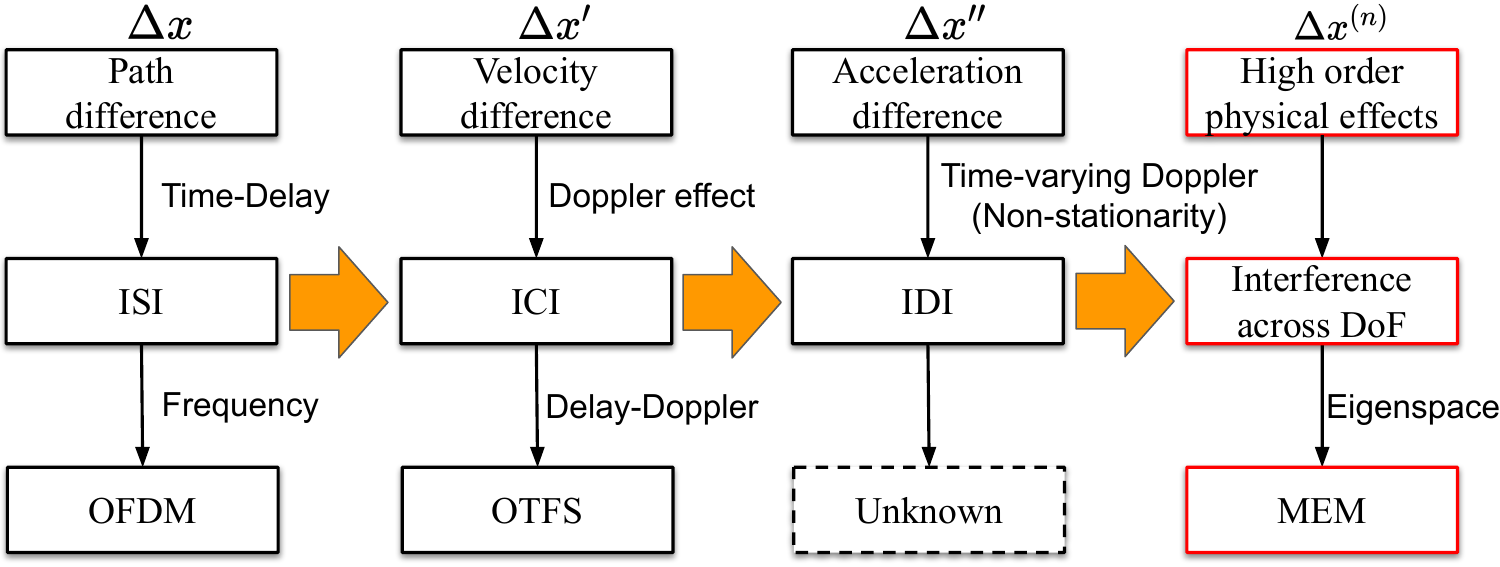}
    \caption{The evolution of conventional modulation techniques}
    \label{fig:evo}
    \vspace{-10pt}
\end{figure}

It is well-known that path delays lead to ISI, which can be mitigated by OFDM by transmitting symbols in the frequency domain. On the other hand, the Doppler effect causes ICI, which can be mitigated by OTFS modulation, where symbols are transmitted in the Delay-Doppler domain. 
However, in NS channels, both the delay and the Doppler effects change over time and frequency, which leads to interference in the delay-Doppler domain as well, referred as IDI~\cite{Raviteja2018eq_otfs}. Although detectors have been proposed to mitigate IDI~\cite{Raviteja2018eq_otfs, thaj2020low}, it cannot ensure interference-free in the delay-Doppler domain. 
As shown in figure~\ref{fig:evo}, these techniques are developed iteratively to mitigate ISI, ICI and then IDI due to \textit{path difference ($\Delta x$), velocity difference ($\Delta x^{\prime}$) and acceleration difference ($\Delta x^{\prime\prime}$)}, by investigating orthogonality in the time, time-frequency and delay-Doppler domain, respectively. In general, modulation uses carriers in the domain represented by high order physics as it is relatively less variant causing minimal interference. However, in practice there may exist higher order physical variation $\Delta x^{(n)}$ under more complex xG communication scenarios, leading to interference across the DoF. This motivates the Multidimensional Eigenwave Multiplexing (MEM) modulation, which designs carriers in the eigendomain by employing {\abbrev} to generate 
jointly orthogonal eigenfunctions (or eigenwaves). These modulated symbols can achieve joint orthogonality while transmitting over NS channels with high-dimensional variations. 

%% file: decomposition.tex
\section{Decomposition of the MU-MIMO NS Channel}
\label{sec:NS}

\subsection{General NS Channel Model}
\label{sec:model}
For SU-SISO NS channels, 
the NS channel is commonly represented using time-varying impulse response $h(t,\tau)$, spreading function $S_H(\tau, \nu)$ or TF transfer function $L_H(t,f)$, depending on the domain used to transmit the signal and observe the channel (also the domain for estimating the CSI). These functions are used to mathematically relate the transmitted and the received signal.
For MU-MIMO NS channels, let's consider time-varying response $h(t,\tau)$ and expand it to incorporate the space domain. Then the channel can be modeled as $\mathbf{H}(t,\tau)$, which is a 4-D channel tensor. Signal transmitted over this 4-D channel will incur spatial and temporal interference from other users and its own delayed symbols, respectively.

Now, if we consider the channel as a continuous system with respect to both space and time dimensions, which maps the space-time signal, $s(u^{\prime},t^{\prime})$ from the transmitter to the receiver by a \textit{Channel Kernel, $k_H(u,t;u^{\prime},t^{\prime})$}. Without loss of generality, we denote $u$ and $u^{\prime}$ as a variable in the space domain that represents both users and antennas. Then received signal is expressed as,
\begin{align}
    \label{eq:MU2}
    & r(u,t) = \iint k_H(u,t;u^{\prime},t^{\prime}) s(u^{\prime},t^{\prime})~du^{\prime}~dt^{\prime} + v(u,t)
\end{align}
where $(u^{\prime},t^{\prime})$ and $(u,t)$ correspond to the space-time domain at the transmitter and receiver, respectively. $v(u,t)$ is the noise. 
$k(u,t;u^{\prime}, t^{\prime}) {=} h_{u,u^{\prime}}(t, t{-}t^{\prime})$ is the channel kernel, where $h_{u,u^{\prime}}(t, \tau)$ is the element of $\mathbf{H}(t,\tau)$ at coordinate $(u, u^{\prime})$. The channel kernel can be understood as a transfer function (different from TF transfer function above) that maps signals from the transmitter to the receiver in the \textit{same domain} (($u,t$ and ($u^{\prime},t^{\prime}$) in this case) though at different ends (Tx/Rx), which is an alternate representation of the channel and can be obtained by the coordinate shift of the channel impulse response $h_{uu^{\prime}}$. Therefore, we find that this joint interference (from both $u^{\prime}$ and $t^{\prime}$) varies simultaneously along the space and time dimensions.
This is referred to as the \textit{dual space-time variation property}. In other words, it implies that joint orthogonal bases in the space-time domain at the transmitter ($(u^{\prime},t^{\prime}$)
is not sufficient to ensure interference-free communication, unless the bases remain orthogonal after propagating through the channel, meaning also orthogonal in the space-time domain at the receiver ($(u,t)$), which is referred as \textit{dual orthogonality}.

\subsection{{\abbrev}: High Order Generalized Mercer's Theorem}
\label{sec:HOGMT}

The central challenge in designing waveforms for MU-MIMO NS channel is the decomposition of the 4-D channel kernel ($k_H$) into 2-D eigenfunctions (as the transmitter or receiver operates in a spatio-temporal 2-D space) that possess the dual orthogonality property mentioned above. While SVD is only capable of decomposing separate channel matrices, Karhunen–Loève Transform (KLT)
provides a method to decompose random processes into component eigenfunctions within the same dimension. 
However, KLT is unable to decompose the 
4-D channel into orthonormal 2-D space-time eigenfunctions (DoF decorrelation) and therefore cannot mitigate joint interference in the space-time dimension.
On the other hand, Mercer's theorem can decompose random processes into eigenfunctions at different dimensions but is only applicable to symmetric processes (function whose value remain the same even after interchanging the variables). However, the NS channel kernel is not necessarily a symmetric process. 
Therefore, a generalized version of Mercer's Theorem for asymmetric processes has been proposed, which extends to high-dimensional cases as well to decompose any asymmetric 4-D channel into 2-D jointly orthogonal eigenfunctions~\cite{ZouICC22}. In {\abbrev}, this is mathematically expressed as, 

\begin{align}
\label{eq:thm2_decomp}
&k_H(u,t;u^{\prime},t^{\prime}) = \sum\nolimits_{n{=1}}^N \sigma_n \psi_n(u,t) \phi_n(u^{\prime},t^{\prime})
\end{align}
where $\psi_n(u,t)$ and $\phi_n(u^{\prime},t^{\prime})$ are eigenfunctions with orthonormal properties. $\sigma_n$ is a zero-mean random variable whose variance are the eigenvalues. The orthonormal property of the eigenfunctions in \eqref{eq:thm2_decomp} ensures that the 4-D kernel is decomposed into jointly orthogonal eigenfunctions (or subchannels). Moreover, decomposed eigenfunctions show desired dual orthogonality as follows, 
\begin{align}
    \iint k_H(u,t;u^{\prime},t^{\prime}) \phi_n^*(u^{\prime}, t^{\prime}) ~du^{\prime} ~dt^{\prime} {=} \sigma_n \psi_n(u, t).
    \label{eq:them2_duality}
\end{align}
This suggests that when $\phi_n$ is transmitted through the 4-D channel, it is transferred to $\psi_n$ with a scaling factor $\sigma_n$, meaning that decomposed dual joint space-time orthogonal subchannels experience a flat-fading. Therefore, {\abbrev} addresses the first part of the open problem posed in Section~\ref{sec:intro}.
We refer to $\phi_n$ and $\psi_n$ as a pair of \textit{dual eigenfunctions}. 

%% file: precoding.tex
\section{{\abbrev}-Precoding for MU-MIMO NS Channels}
\label{sec:Pre}
\begin{figure}
    \centering
    \includegraphics[width=1\linewidth]{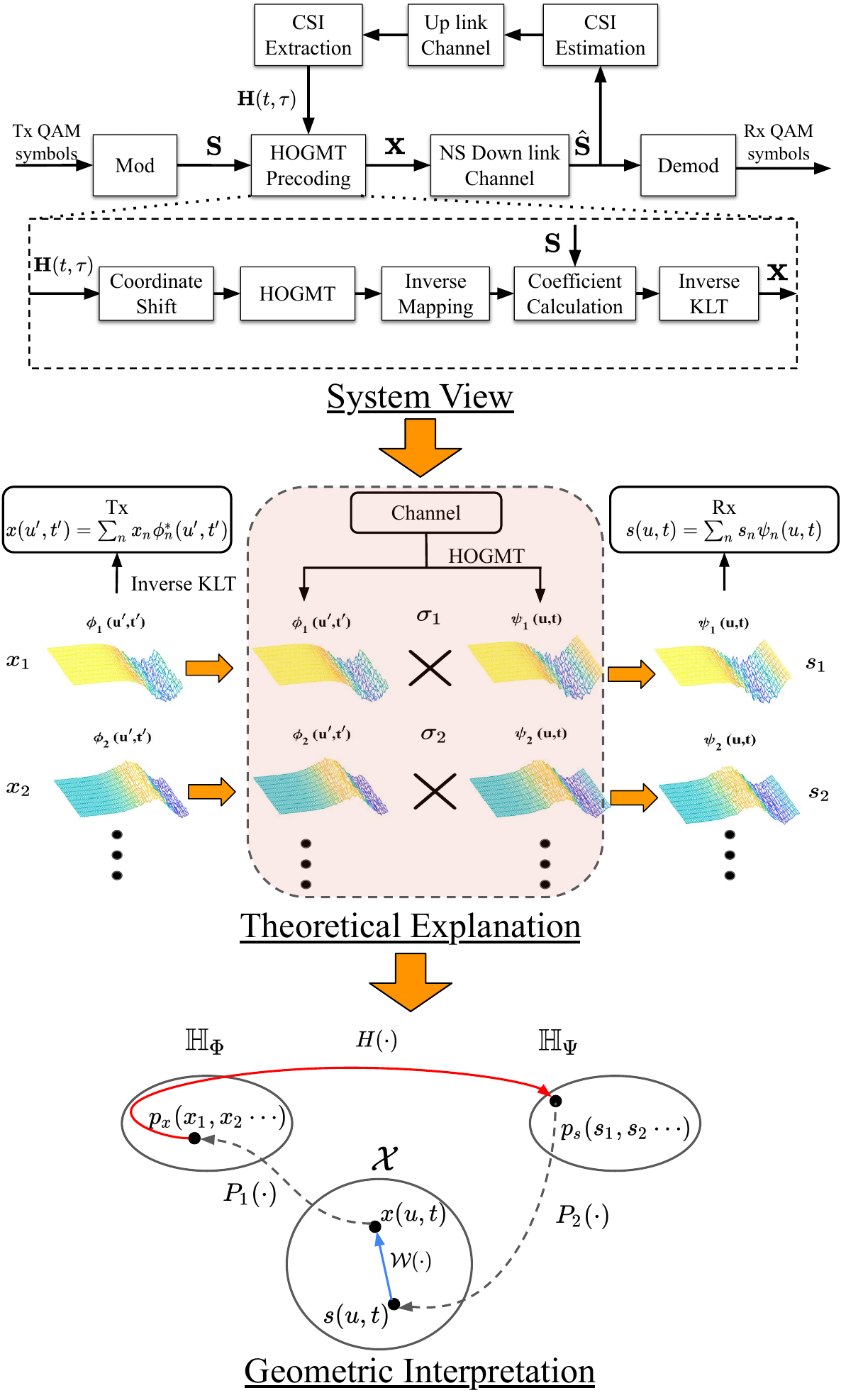}
    \caption{{\abbrev}-Precoding from different points-of-view}
    \label{fig:H_P}
    \vspace{-10pt}
\end{figure}

At a high-level, {\abbrev} decomposes the space-time domain of the transceiver into two uncorrelated eigendomains. Then the transmitter and receiver can be seen as \textit{speaking two different languages} at these two eigendomains. {\abbrev}-Precoding acts as the ``Translator", which helps the transmitter to communicate in a certain way so that the receivers as ``Listeners" can correctly understand the desired meaning without any errors. The CSI is equivalent to a common ``Dictionary", whose accuracy ultimately determines the performance of the Translator. With perfect CSI, {\abbrev}-Precoding can help the transceiver communicate without any potential barrier (in form of interference). Figure~\ref{fig:H_P} illustrates {\abbrev}-Precoding from a system view, theoretical underpinning and geometric interpretation as discussed below. 

\noindent \underline{\textbf{System view:}} The system diagram shows a suggested implementation of {\abbrev}-Precoding. The spatio-temporal CSI obtained from the receivers is used to extract the 4-D channel kernel (an important future research direction), which can be decomposed into dual 2-D eigenfunctions using 
{\abbrev} to decorrelate the space-time domains at the transmitter and the receiver.  The 2-D eigenfunctions corresponding to the receiver are used to derive optimal coefficients that minimize the least square error in the transmitted and received symbols. Then combining their dual spatio-temporal eigenfunctions with these coefficients via inverse KLT. Since the eigenfunctions are jointly orthonormal sub-channels over space and time, 
precoding using these warrants flat-fading (interference-free communication) even in the presence of joint space-time interference. Further, these precoded symbols directly reconstruct the data symbols at the receiver when combined with calculated coefficients. Therefore, unlike existing methods that require complementary decoding at the receiver~\cite{Cho2010MIMObook}, {\abbrev}-Precoding eliminates any receiver processing, significantly reducing the computational burden or the transceiver. 

\noindent
\underline{\textbf{Theoretical Explanation:}} The eigendomain expression shows the theory behind {\abbrev}-Precoding. To mitigate the dual space-time variation property of the channel kernel, we have to find the bases that have dual orthogonality as mentioned in section~\ref{sec:model}, which is conveniently provided by {\abbrev} discussed in Section~\ref{sec:HOGMT}.
If $\Phi$ and $\Psi$ is a pair of dual eigenfunction sets, then transmitting the set $\Phi$ over the channel, a receiver can obtain its dual eigenfunction set $\Psi$ according to \eqref{eq:them2_duality}. Therefore, to reconstruct the desired data signal using eigenfunction set $\Psi$ at the receiver, we can first project data signal $s(u,t)$ onto $\Psi$ to obtain projections $\{s_n\}$. Then the coefficients of $\Phi$, $\{x_n\}$ can be obtained by dividing $\{s_n\}$ by $\{\sigma_n\}$. Combing eigenfunctions in $\Phi$ with corresponding coefficients $\{x_n\}$, we have the precoded signal, $x(u,t)=\sum_n x_n \psi_n^* (u,t)$, which answers the second part of the open problem statement in Section~\ref{sec:intro} about signalling scheme for NS channels.

Therefore, {\abbrev}-Precoding can be seen as transmitting the eigenfunction set $\Phi$ after multiplying with derived coefficients $\{x_n\}$, which will transfer to $\{s_n\}$ after scaled by $\{\sigma_n\}$.
Then the data signal $s(u,t)$ is directly reconstructed at the receiver by the dual eigenfunction set $\Psi$ with transferred coefficients $\{s_n\}$ ensures that $r(u,t){=}\sum_n s_n \psi_n(u,t){+}v_n(u, t){=} s(u,t) + v(u,t) {=} \hat{s}(u,t)$, 
where $\hat{s}(u,t)$ is the estimated signal. 
Therefore, the {\abbrev} decomposition of the channel allows us to precode the signal such that all interference in the space domain, time domain and joint space-time domain are canceled when transmitted through the channel, leading to a joint spatio-temporal precoding scheme. 
Further, this precoding ensures that the modulated symbol is reconstructed directly at the receiver with an estimation error that 
of $v(u, t)$, 
thereby completely pre-compensating the spatio-temporal fading/ interference in NS channels to the level of AWGN noise.
Therefore, this precoding does not require a complementary step at the receiver, which vastly reduces its hardware and computational complexity compared to state-of-the-art precoding methods like DPC or SVD precoding.

\noindent
\underline{\textbf{Geometric interpretation:}} Given two Hilbert Space $\mathbb{H}_\Phi$ and $\mathbb{H}_\Psi$, with bases as the eigenfunction sets $\Phi$ and $\Psi$, respectively, the precoded signal $x(u,t){\in} \mathcal{X}$, where $\mathcal{X}$ is the space of reality (i.e., space-time domain of transceivers), can be seen as a point $p_x \in \mathbb{H}_\Phi$, where the coordinate of this point is the coefficient (power) allocated to the corresponding eigenfunction. Then the 4-D channel kernel has the mapping $H(\cdot): p_x \to p_s$, where the point $p_s \in \mathbb{H}_\Psi$ represented in space of reality, i.e., projected to  $\mathcal{X}$ is directly the data signal $s(u,t) \in \mathcal{X}$. The dual spatio-temporal variation of the 4-D channel not only transfers the coordinate of $p_x$ to $p_s$, but also transfer Hilbert Space $\mathbb{H}_\Phi$ to $\mathbb{H}_\Psi$. As {\abbrev} extracts the duality, which explains the transformation $H(\cdot)$, and dual orthogonality, which provides the complete projections $P_1(\cdot)$ and $P_2(\cdot)$, we can use the inverse method to construct the precoded signal, i.e., for the target closed loop $\mathcal{W}(\cdot) + P_1(\cdot) + H(\cdot) + P_2(\cdot) = 0$, we have $\mathcal{W}(\cdot) = - P_2(\cdot) - H(\cdot) - P_1(\cdot)$, meaning, $\mathcal{W}(\cdot)$ can be obtained by the inverse process $s(u,t) \to p_s \to p_x \to x(u,t)$, which is analogous to the theoretical explanation of {\abbrev}-Precoding in previous sections. 

{\abbrev}-Precoding is evaluated using the 3GPP 38.901 UMa NLOS scenario built on QuaDriga in Matlab. The base station (BS) is equipped with $10$ antennas. There are $5$ user equipments (UEs) and each of them is equipped with $2$ antennas. The speed of UEs is $120 \pm 18$ km/h. 
Figure \ref{fig:ber_space_time} shows the 
BER at the receiver with {\abbrev}-precoding with 16-QAM modulated symbols. Since this precoding is able to cancel all space-time varying interference that occurs in space, time and across space-time dimensions, it achieves significantly lower BER compared to DPC that is applicable for the time-invariant channels. 
Further, we show that with more eigenfunctions we can achieve lower BER. With more than $99\%$ eigenfunctions, it can achieve near-ideal BER, where the ideal case assumes all interference is canceled and only AWGN noise remains at the receiver. Although using more eigenfunctions achieves lower BER, it will cost more energy. The trade-off between energy and interference cancellation is associated with the maximum energy efficiency criteria, which is discussed as one of the future research directions in Section~\ref{sec:fut}.

%% file: modulation.tex
\section{MEM Modulation for MU-MIMO NS Channels}
\label{sec:Mod}

The recently proposed OTFS modulation design carriers in the delay-Doppler domain achieves orthogonality in the time-frequency domain but fail to maintain orthogonality in MU-MIMO NS channels because of mutual correlation between the Delay-Doppler and TF domains. This motivates further investigation into identifying symbol carriers to achieve orthogonality across all DoF.

\noindent{\underline{\textbf{Theoretical overview:}} So far, we have seen that {\abbrev} can decompose a 4-D channel into eigenfunctions that are jointly orthogonal across its DoF. Different from {\abbrev}-Precoding, which projects the entire data signal $s(u,t)$ onto eigenfunction set $\Psi$, MEM uses these as data-carriers by modulating each eigenfunction with the data symbol $s_n$ as $s_n \psi^*_n$. Since the eigenfunctions are represented as a waveform at its DoF in MEM, it is convenient to refer to as eigenwaves in the context of modulation.  Note that $s_n$ in MEM modulation is a QAM symbol instead of projections as in {\abbrev}-Precoding in Section~\ref{sec:Pre}. Multiplexing all symbols and carriers, we have the transmitted signal, $x {=} \sum_n s_n \phi_n^*$. Due to duality, after transmission over the channel, $x$ is transferred to $r {=} \sum_n \sigma_n s_n \psi_n + v_n$ at the receiver, where  $v_n$ is the noise. As eigenfunctions have the orthonormal property, $r$ is demodulated by a Matched Filter using each $\psi_n^*$. Then the estimated symbol is obtained as $\hat{s}_n {=} \sigma_n s_n + v_n$, which suggests that the demodulated symbol $\hat{s}_n$ is the data symbol $s_n$ multiplied a scaling factor (channel gain) $\sigma_n$ along with AWGN, meaning there is no interference from other symbols.

\begin{figure}
    \centering
    \includegraphics[width=\linewidth]{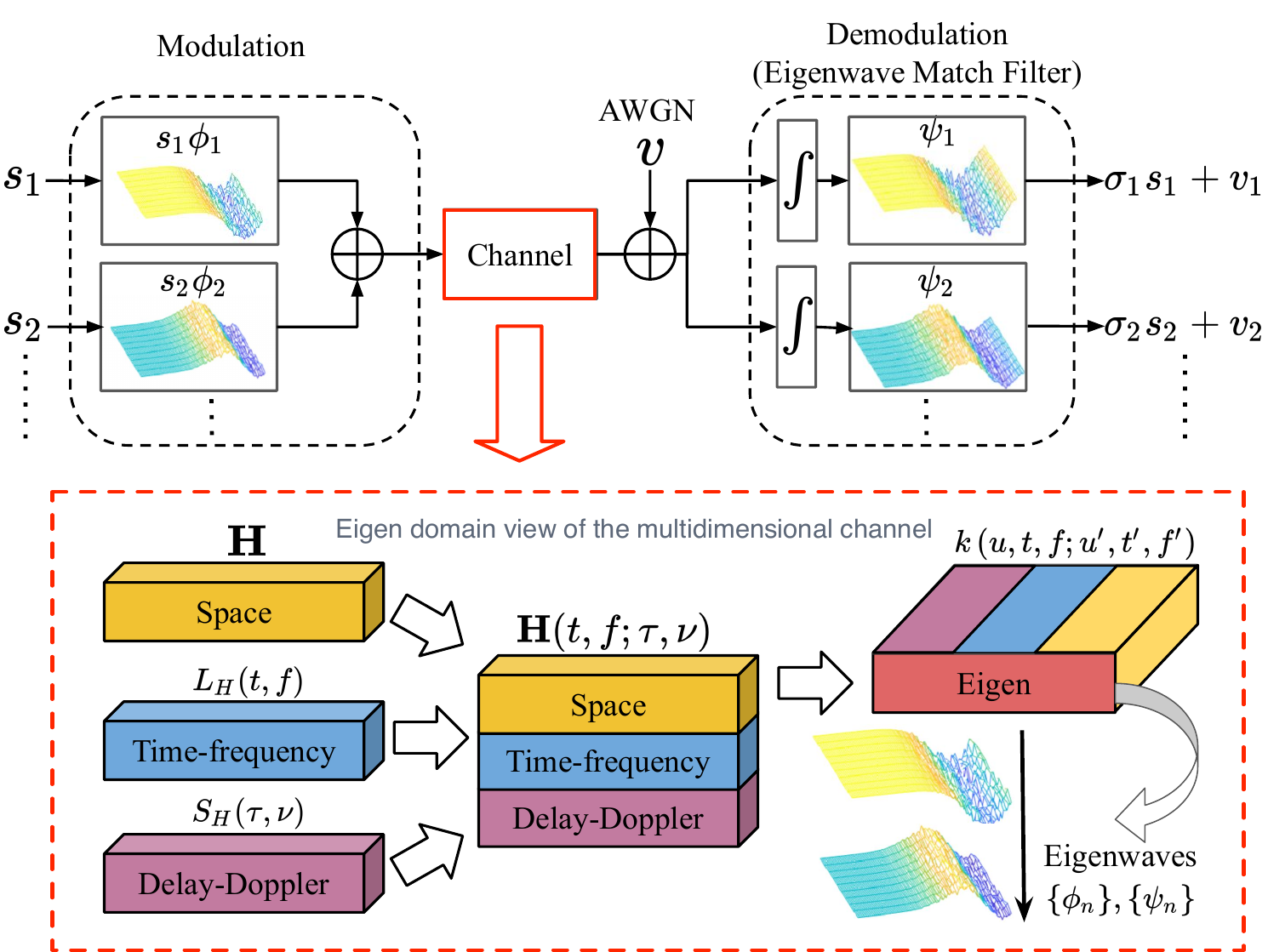}
    \caption{Multidimensional Eigenwave Multiplexing}
    \label{fig:2-D example}
    \vspace{-10pt}
\end{figure}

\begin{figure*}
\centering
\begin{subfigure}{.295\textwidth}
\centering
  \includegraphics[width=1\linewidth]{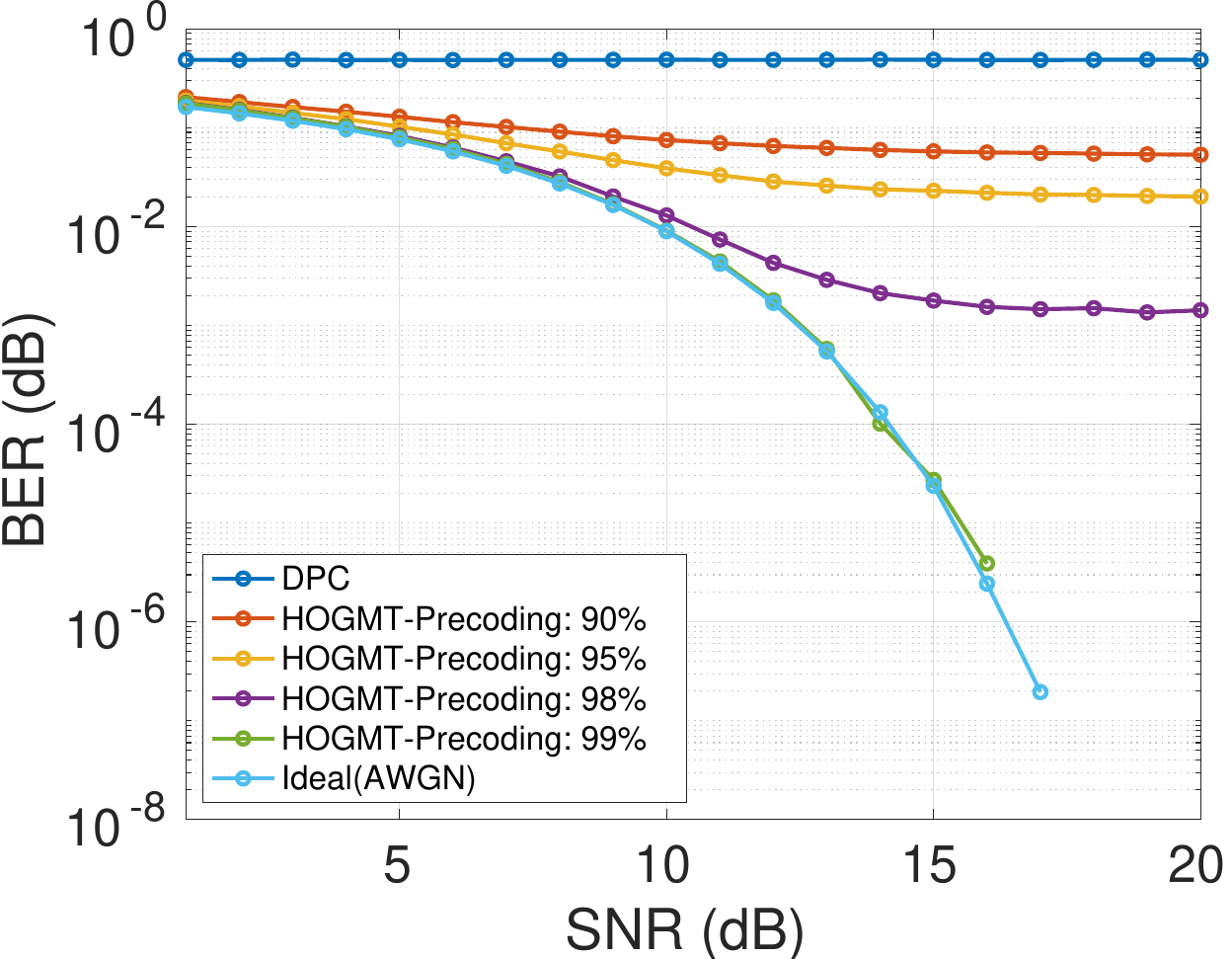}
  \caption{BER of {\abbrev}-Precoding}
  \label{fig:ber_space_time}
\end{subfigure}
\qquad
\begin{subfigure}{.3\textwidth}
  \centering
  \includegraphics[width=1\linewidth]{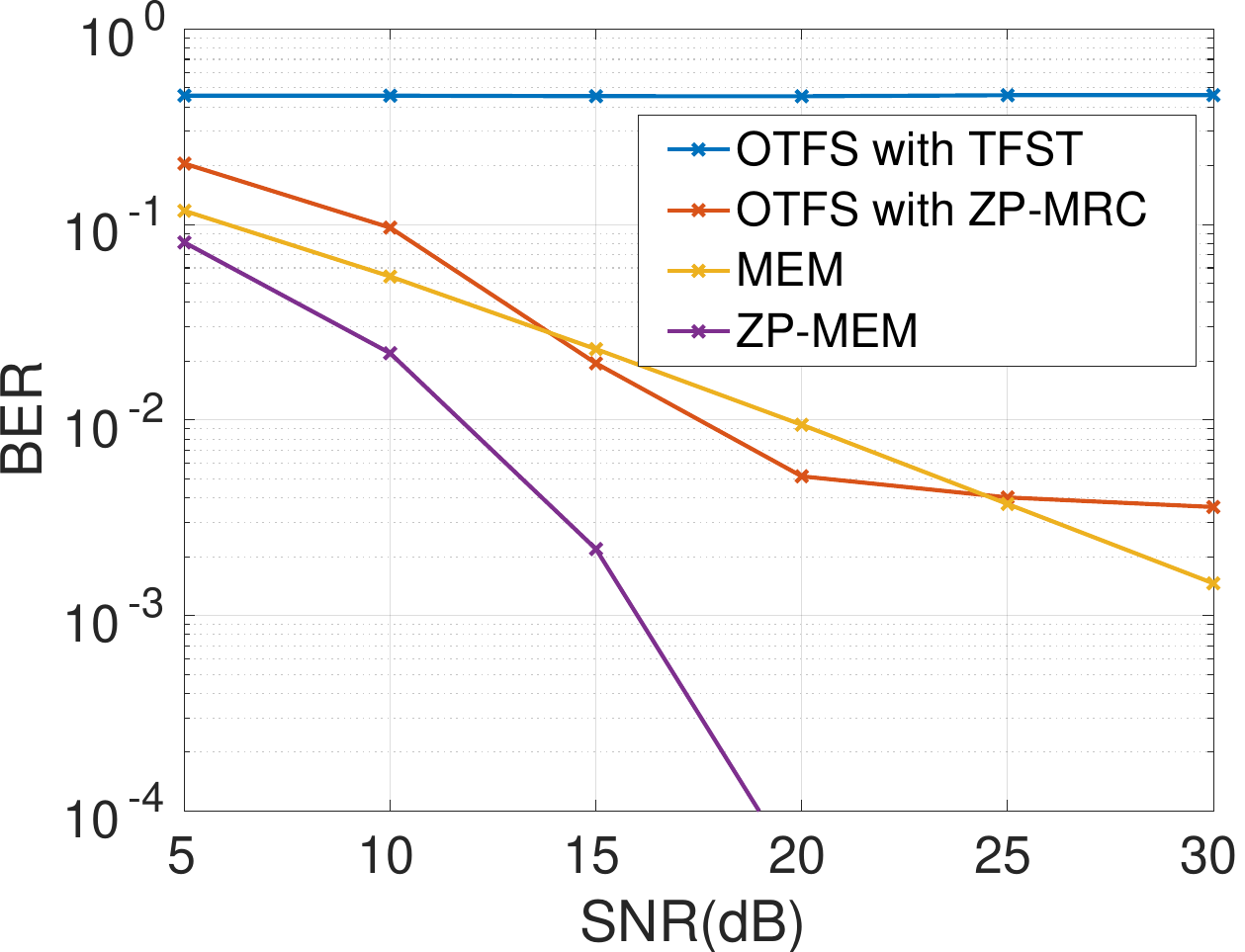}
  \caption{BER of MEM}
  \label{fig:ber_b}
\end{subfigure}
\qquad
\begin{subfigure}{.295\textwidth}
  \centering
  \includegraphics[width=1\linewidth]{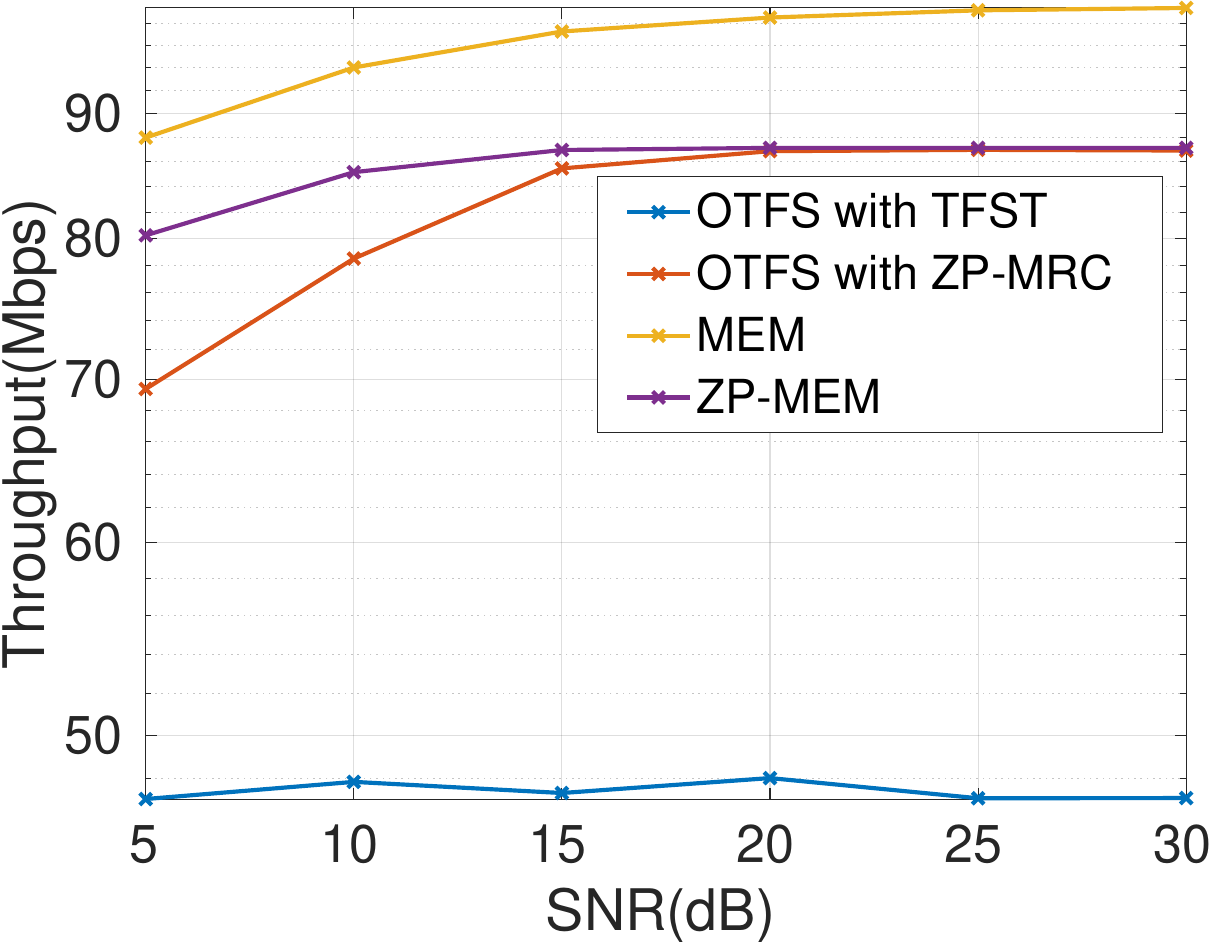}
 \caption{Throughput of MEM}
  \label{fig:th_b}
\end{subfigure}
  \caption{Results of {\abbrev}-Precoding and MEM modulation and its comparison with SoTA}
  \label{fig:cb}
  \vspace{-10pt}
\end{figure*}
Figure~\ref{fig:2-D example} shows an example of eigenwave modulation using {\abbrev}. At the transmitter, each data symbol, $s_n$ is multiplied by one eigenwave $\phi_n$, obtained by {\abbrev} decomposition, and then summed to create the modulated signal. The data symbols remain independent during transmission over the channel due to the joint orthogonality of eigenwaves. At the receiver, each data symbol estimate, $\hat{s}_n$ is obtained by a matching filter using the eigenwave $\psi_n$, also obtained as a product of {\abbrev} decomposition, giving the data symbol $s_n$ multiplied by the corresponding channel gain, $\sigma_n$ with AWGN. 

This implies that the data symbol, $\{s_n\}$ has leveraged all the diversity gain. The reason is that the diversity of the multidimensional channel at each DoF (space, time-frequency, delay-Doppler) are merged (integral along each DoF) and then divided in the eigendomain into independently flat-fading subchannels or eigenwaves as shown in Figure~\ref{fig:2-D example}. Therefore, ({\abbrev} is a generalized decomposition method for the multidimensional processes, and thus~\eqref{eq:thm2_decomp} can be directly extended to 6-D channel kernels as well. For example, the data signal in \eqref{eq:MU2} can also represent a 3-D space-time-frequency domain as $s(u, t, f)$. In this case, the channel kernel is extended to 6-D, which maps $(u,t,f)$ to $(u^{\prime},t^{\prime},f^{\prime})$)}. 
Furthermore, since eigenwaves achieve diversity in eigenspace, it can be inferred as \textit{diversity achieving} for the total channel as well, since the eigenspace is simply an alternate view of the multi-dimensional NS channel.


\noindent{\underline{\textbf{Evaluation:}} 
MEM for NS channels is evaluated in Matlab using the Extended Vehicular A (EVA) model. The speed range is $[100, 150]$ km/h and the stationarity interval is one OFDM subcarrier. We also present comparisons to OTFS with time-frequency single tap (TFST)~\cite{Hong2022DD} detector and Zero-Padded maximal ratio
combining (ZP-MRC)~\cite{thaj2020low} detector. For a fair comparison, we also implement a Zero-Padded MEM (ZP-MEM) version, where \textit{zero pad} is placed on eigenfunctions with least $\sigma_n$. ZP length is $1/8$ symbol for both ZP-MEM and ZP-MRC. 

Figure~\ref{fig:ber_b} compares the BER of MEM, ZP-MEM, OTFS with TFST and OTFS with ZP-MRC. OTFS with TFST detector does not work at all in this case and OTFS with ZP-MRC detector has a similar BER as MEM. This is because demodulating data symbols on carriers (eigenwaves) with the least $\sigma_n$ will enhance the noise as well. ZP-MEM doesn't put data symbols on these eigenwaves, thereby achieving lower BER. However, for both OTFS with ZP-MRC and ZP-MEM, the throughput is less than MEM as the zero pad increases overhead, which is shown in figure~\ref{fig:th_b}. However, optimal utilization of MEM carriers is certainly a promising future research direction.


%% file: futurework.tex
\section{Qualitative Comparison between HOGMT-Precoding and MEM modulation}
\label{sec:comp}
{\abbrev}-Precoding and MEM modulation use the same mathematical tool {\abbrev} and both leverage dual eigenfuntions to achieve the orthogonality. The main difference is that HOGMT-Precoding uses eigenfunctions to deconstruct and reconstruct the whole data signal at the transmitter and receiver, respectively. The eigenfunctions are used as projection bases in this case, the number of which would affect the completeness of the projection. Therefore, insufficient eigenfunctions lead to the reconstruction error at the receiver. The coefficient of the eigenfunction set $\Psi$ at the transmitter $\{x_n\}$ is the energy cost. Although using more eigenfunctions can achieve lower BER, it increases the energy cost. While for MEM modulation, each eigenfunction is used as a data-carrier. Therefore, the coefficient of $\Psi$ is not the allocated energy but the data symbol $\{s_n\}$. Carried by these eigenfunctions, data symbols can parallelly transmit over orthogonal subchannels. Then the estimation of each data symbol is independent of other symbols and eigenfunctions. In this case, the number of eigenfunctions only affects the throughput. However, symbols carried by eigenfunctions with small channel gains would enhance the noise after demodulating, which leads to estimation error. 

\section{Future Research Directions}
\label{sec:fut}
In the above sections, we have discussed the decomposition, precoding and modulation for non-stationary channels. These techniques are optimal with respect to interference cancellation, though require high accuracy of CSI or high energy cost. 
In general, there is always a trade-off between optimality and robustness. With the imperfect CSI, the performance of discussed techniques will certainly degrade but we believe that robust techniques can be developed based on the reviewed techniques in this article, which provide the theory support and benchmark for further studying. 

\subsection{Spatio-temporal CSI}
The most relevant 4-D CSI can be obtained by the double-selective channel sounder, but it may not provide the CSI in a timely manner for certain applications. There are some investigations on 4-D CSI estimation in the literature, such as~\cite{2020CESrivastava}. 
However, in a simpler setting, where the delay effect does not change in the space domain (users/antennas), the 4-D CSI $\mathbf{H}(t,\tau)$ can be obtained by the Kronecker product of the impulse response and the space channel matrix as $\mathbf{H}\otimes h(t,\tau)$. Both of these can be obtained using contemporary methods.

\subsection{Robustness with imperfect CSI}
\noindent
Different from MEM modulation, HOGMT-Precoding projects the whole data signal onto the decomposed eigenfunctions. Therefore, once there is a CSI error, which leads to the eigenfunction error, the whole data signal will undergo a reconstruction error at the transceiver. If we can evaluate this error caused by each eigenfunction, even in a statistical manner, and compare it with the canceled interference of this eigenfunction. Then there exists a trade-off on whether use the eigenfunction as a part of bases or not. While for the MEM, the error in each eigenfunction only affects the estimation of the data symbol carried by this eigenfunction. Then the trade-off of whether use one eigenfunction as a data-carrier is determined by if the carried data symbol can be estimated with the desired accuracy. 

\subsection{Energy efficiency}

With respect to the maximum energy efficiency, the more eigenfunctions used in HOGMT-Precoding, the more interference can be canceled. The interference cancellation is basically linear with respect to the number of eigenfunctions used. That's because the current linear implementation of {\abbrev}  only decomposes eigenfunctions for approximating the channel kernel. Eigenfunctions with more eigenvalue extract more information of the channel kernel, however, are just orthogonal bases with basically equal contributions for reconstructing the data signal. Theoretically, there exists an optimal nonlinear implementation using tools such as deep learning, that the decomposed eigenfunctions have high eigenvalues (projections) with respect to both channel kernels and data signals. Then we can choose the eigenfunctions with the most eigenvalues to achieve maximum energy efficiency. 

For MEM modulation, data-carriers have different channel gains. Symbols on the data-carriers with the least gains would incur the most noise enhancements during demodulation. The zero-padding MEM just simply discards these data-carriers. 
Choosing appropriate carriers and allocating the power for MEM modulation by certain criteria, like the water-filling algorithm, is worth further studying.

%% file: conclusion.tex
\section{Conclusion}
\label{sec:conclusion}
This article shows the evolution and limitations of waveform design techniques for MU-MIMO channels. Based on that we give an overview of {\abbrev} decomposition, which is able to decompose any high-dimensional, time-varying process into jointly orthogonal eigenfunctions across its DoF. Empowered by this method, we discuss its applicability for designing waveforms for xG non-stationary channels using two approaches: 1) {\abbrev}-Precoding, that uses the decomposed eigenfunctions to deconstruct and reconstruct the signal; 2) MEM modulation that employ the same eigenfunctions as carriers. Both methods perform better in terms of BER compared to the SoTA for MU-MIMO NS channels. In spite of promising results, we believe that there are a number of interesting open problems that stem from this discussion as well as future improvements, opening a broad and exciting new research area for the years to come.
  

%% file: ack.tex
\section{Acknowledgement}
This work is supported by the NSF CAREER project \#2144980 and the Air Force Research Laboratory Visiting Faculty Research Program (SA10032021050367), Rome, New York, USA.

%% file: bio.tex
\section{Biographies}
\noindent
\textbf{Zhibin Zou} is an PhD student in the Department of Electrical and Computer Engineering at University at Albany, SUNY. He received his Master's degrees from Xidian University, China. His research focuses on wireless communications, precoding and channel characterization with an emphasis on waveform design for non-stationary channels using ML/DL techniques. He is a recipient of the \textit{Best Paper Award} within the \textit{Wireless Communication track} in IEEE ICC 2022.
\vspace{3mm}

\noindent
\textbf{Aveek Dutta} is currently an Assistant Professor in the Department of Electrical and Computer Engineering at University at Albany, SUNY. Previously, he was an Assistant Professor at University of Kansas and Postdoctoral Research Associate at Princeton University and has PhD and MS degrees in Electrical Engineering from University of Colorado Boulder. His research spans across various topics in the areas of generalization of deep learning for wireless communication, distributed and trustworthy spectrum sensing and policy enforcement, Heterogeneous signal processing in high-speed Cloud-RAN and applied machine learning for interference cancellation in radio astronomy. In recent years he has been the recipient of best paper awards in IEEE ICC and DySPAN conferences.